# Quantum key distribution over a 72 dB channel loss using ultralow dark count superconducting single-photon detectors

Hiroyuki Shibata,[1,2,*] Toshimori Honjo,[3] and Kaoru Shimizu[1]

[1]NTT Basic Research Laboratories, NTT Corporation, 3-1 Morinosato-Wakamiya, Atsugi, Kanagawa 243-0198, Japan
[2]NTT Nanophotonics Center, NTT Corporation, 3-1 Morinosato-Wakamiya, Atsugi, Kanagawa 243-0198, Japan
[3]NTT Secure Platform Laboratories, NTT Corporation, 3-9-11 Midori-cho, Musashino, Tokyo 180-8585, Japan
*Corresponding author: shibata.h@lab.ntt.co.jp

We report the first Quantum key distribution (QKD) experiment over a 72 dB channel loss using superconducting nanowire single-photon detectors (SSPD, SNSPD) with the dark count rate (DCR) of 0.01 cps. The DCR of the SSPD, which is dominated by the blackbody radiation at room temperature, is blocked by introducing cold optical bandpass filter. We employ the differential phase shift QKD (DPS-QKD) scheme with a 1 GHz system clock rate. The quantum bit error rate (QBER) below 3 % is achieved when the length of the dispersion shifted fiber (DSF) is 336 km (72 dB loss), which is low enough to generate secure keys.

Quantum key distribution (QKD) enables two remote participants to share ultimately secure keys based on the laws of quantum mechanics [1,2]. Since the first demonstration of a QKD experiment using a 32-cm free-space transmission in 1992 [3], the achievable distance for QKD has improved significantly [4-8], and QKD over 260 km (52.9 dB loss) has been reported [4]. As the channel loss increases in QKD, the number of signals that reach Bob decreases, and the quantum bit error rate (QBER) deteriorates because the dark counts dominate in the signals. We cannot generate a secure key from the sifted key when QBER exceeds a certain level which depends on QKD schemes. This limits the QKD distance that can be achieved without resorting to complicated quantum relays and repeaters [9]. Therefore, we need a single-photon detector with a low dark count rate (DCR) at telecom wavelengths for long-distance QKD. There have already been many studies employing long-distance QKD experiments with different types of single-photon detectors, such as an InGaAs avalanche photodiode (InGaAs-APD), a frequency-upconversion assisted silicon APD, and a superconducting nanowire single-photon detector (SSPD or SNSPD) [4-8,10-12]. Of these, the SSPD has the lowest DCR of $10^{-4}$ to 1000 cps, which depends on the bias current and its temperature [13-17], and it has been extensively used. Takesue et al. performed a QKD experiment with a channel loss of over 42.1 dB using SSPDs with a DCR of 50 cps and a system detection efficiency ($\eta$) of 1.4 % [6]. Recently, a 52.9 dB channel loss was achieved by Wang et al. using SSPDs with a DCR of 1 cps and an $\eta$ value of 3 % by greatly reducing the bias current of the SSPDs [4]. A further increase in the maximum channel loss of QKD experiments requires a single-photon detector with a higher $\eta$/DCR ratio.

Recently, we have developed a SSPD with ultralow DCR (<0.1 cps) by introducing optical bandpass filters and cooled them at low temperature [16]. The cold filter removes the DCR from the blackbody radiation at room temperature which propagate through optical fiber. Figure 1 shows the system $\eta$ and system DCR of the SSPD. At bias current of 21.5 μA, $\eta$ = 4.4 % and DCR < 0.01 cps is achieved at 0.38 K by introducing the cold optical filter with the bandwidth of 20 nm and blocking down to 3 μm. Here we report a QKD experiment exceeding 72 dB channel loss using the SSPD with ultralow DCR.

We employed a differential phase shift QKD (DPS-QKD) scheme using a weak coherent light [18,19], which is suitable for stable, high-speed and long-distance QKD experiments with a simple setup. Figure 2 shows a schematic diagram of our experimental setup. At the transmitter's (Alice's) site, a continuous wave laser (Agilent Technologies, 81600B) emitting at $\lambda$ = 1547 nm is modulated into a 1 GHz pulse train with a 100 ps pulse width by using an intensity modulator (IM, Sumitomo Osaka Cement, T.MXH1.5). Then the phase of each pulse is randomly modulated by 0 or $\pi$ using a phase modulator (PM). The pulse train is strongly attenuated to 0.2 photon/pulse by an optical attenuator and then sent to the quantum transmission line. For the quantum transmission line, we employ dispersion-shifted fibers (DSF) with lengths of 306 km (66 dB channel loss) and 336 km (72 dB loss), and an optical attenuator to emulate other channel losses. At the receiver's (Bob's) site, the pulse train is introduced into a 1-bit delay Mach-Zehnder interferometer based on a planar lightwave circuit (MZI-PLC, NTT Electronics, custom-made) and the two outputs of the MZI-PLC are directed to the SSPD system. The SSPD device 1 (D1) clicks when the phase difference between two adjacent pulses is 0, and D2 clicks when the phase difference is $\pi$. Then Bob sends Alice the time stamps of the click events using a public channel. Since both Bob and Alice know the phase difference at that time, they can share a key bit. In DPS-QKD with an average photon number μ of 0.2 photon/pulse, a key that is secure against general individual attacks can be distilled from the sifted key when the QBER is below 4.1 % [19].

Here we specify the security status of current DPS-QKD scheme. For an ideal single-photon emitting source, unconditional security has been proven [20]. However, for the case in which a weak coherent light pulse is used just as in our experimental system, unconditional security has not been proven yet and remains as an open question. Tamaki et al. report unconditional security of a modified DPS-QKD with weak coherent light [21]. However, it needs modification of the system to use block-wise phase randomization and photon number resolving detectors. In this paper, we define the secure key as the final key which is proven to be secure against general individual attacks [19], which is comparable to the previous long-distance QKD experiments [4-6]. We also employ the assumption of infinite key length when we evaluate the bit error rate for the quantum channel. Although the present system does not cope with the sequential attack and the tailored bright illumination attack for QKD, it is possible to upgrade our system to cope with these attacks [22].

The probability of a click event occurring at Bob's site is calculated as follows [19],

$$P_{click} = P_{signal} + P_{dark}$$
$$= \mu \cdot \eta \cdot 10^{-\frac{Loss+Loss_{system}}{10}} + 2 \cdot DCR \cdot t_W \quad (1)$$

where $P_{signal}$ and $P_{dark}$ are the probability of a click event caused by a signal and by the dark count, respectively, $\mu$ is the average photon number, $Loss$ is the channel loss, $Loss_{system}$ is the loss of the MZI-PCL, and $t_w$ is the time window. The sifted key generation rate and QBER are given by,

$$R_{sifted} = \nu \cdot P_{click} \cdot e^{-\nu \cdot P_{click} \cdot t_d} \quad (2)$$

$$QBER = \frac{e_s \cdot P_{signal} + \frac{1}{2} \cdot P_{dark}}{P_{click}} \quad (3)$$

where $\nu$ is the system clock frequency, $t_d$ is the dead time of the TIA, and $e_s$ is the baseline error rate. Finally, the secure key generation rate after error correction and privacy amplification against general individual attack is given by,

$$R_{secure} = R_{sifted} \cdot \left[ -(1-2\mu)log_2\left(1 - QBER^2 - \frac{(1-6 \cdot QBER)^2}{2}\right) + f(QBER) \cdot h(QBER) \right] \quad (4)$$

where $h(QBER) = QBER \cdot log_2(QBER) + (1-QBER) \cdot log_2(1-QBER)$ is the binary Shannon entropy and $f(QBER)$ depends on the error correction algorithm. For the fittings, we take $f(QBER) = 1.2$, $t_d = 20$ ns, $e_s = 0.01$, $\nu = 1$ GHz, $t_w = 100$ ps, $Loss_{system} = 2$ dB, $\eta = 6.7$ and 4.0 % for DCR = 0.04 cps, $\eta = 4.4$ and 3.1 % for DCR = 0.01 cps, and a 50 % reduction in $\eta$ due to the 100 ps time windows.

The DPS-QKD system has been used in a field test in the Tokyo QKD network for a 90 km (27 dB loss) fiber transmission [11]. We measure the sifted key rate and QBER using the system, and secure key rate was calculated from Eq. (4). The extinction ratio and loss of the MZI-PLC were 20 dB and 2 dB, respectively. The extinction ratio limited the QBER of the system to 1 %. The timing of the pulse arrival between Alice and Bob is synchronized by another DSF of the same length as the quantum transmission line to compensate for the timing delay that results from using a long fiber (not shown in Fig. 2). To cope with the fluctuation in the pulse timing caused by the fiber length fluctuation during the QKD experiment, the temperature of the system was stabilized by blocking off an air conditioning of the room. The length of the sifted key was about 250 bits for a 306 km DSF and about 50 bits for a 336 km DSF. We averaged the results 3 times for the 306 km DSF and 6 times for the 336 km DSF.

The experimental results for QKD with SSPDs with a DCR of 0.04 cps are summarized in Fig. 3(a) and in Table 1. The open and closed circles show experimental results for the sifted key generation rate obtained using an optical attenuator and DSF, respectively. The experimental QBER results are also shown by triangles. The square symbols are calculated results for the secure key generation rate obtained using the experimental results for the sifted key rate and QBER based on the theory [19]. The curves are the results of theoretical fitting. As the channel loss increases, the sifted key rate decreases exponentially. The QBER is about 1 % when the channel loss is below 60 dB, indicating that the QBER is determined by the extinction ratio of the MZI-PLC. At a channel loss of 52.7 dB without including PLC-MZI loss, the sifted key generation rate is 31.95 bit/s and the QBER is 1.02 %, resulting in a secure key rate of 12.5 bit/s. The secure key rate is about six times higher than that for the previous longest experimental result obtained with a 260 km (52.9 dB loss) standard telecom fiber [4]. As the channel loss increases to 66 dB (306 km DSF), the sifted key rate and the QBER become 0.98 bit/s and 2.64 %, respectively. The QBER is low enough to generate a secure key with a rate of 0.17 bit/s. Although the theoretical curve indicates that it is possible to generate a secure key even at a channel loss of 72 dB (336 km DSF), we cannot generate a secure key at that length because the QBER exceeds 4.1 %. We attribute this to experimental imperfections such as a slight increase in the DCR caused by stray light and the relatively large jitter of the electronics.

Figure 3(b) summarizes the results of a QKD experiment with DCR = 0.01 cps, when the bias current of the SSPDs is reduced. In this case, the QBER is low as the channel loss is less than around 70 dB. At a channel loss of 72 dB, the sifted key rate and QBER are 0.22 bit/s and 2.93 %, respectively, resulting in a secure key rate of 0.03 bit/s. To approach the asymptotic limit of $10^6$ bits sifted key length, it takes about 2 month, which may be possible by upgrading the system with long-time fluctuation compensation circuit. As the channel loss increases to 77.9 dB, the QBER increases to 10.9 %, so, it is impossible to generate secure keys.

There have already been several QKD experiments undertaken in a field environment [8,10,11]. However, they were all performed in metropolitan areas and the distances were limited to about 100 km. The present results demonstrate the possible application of QKD between two metropolises such as New York and Washington D.C. (328 km) or London and Brussels (320 km), where there is more demand for QKD. If we use ultralow loss fiber (0.164 dB/km) instead of DSF [5], we can further extend the distance to 439 km, which is longer than the direct distances between London and Paris (345 km), Tokyo and Osaka (397 km), or Paris and Geneva (411 km). Although the secure key rate of the present experiment is too small to generate secure keys for a real system, there are several ways to improve the key rate, such as by adopting a 10 GHz clock rate [6], using a high η (>50 %) SSPD device [23-25], and expanding the approach to wavelength-division multiplexing QKD [11]. With these new technologies, it may be possible to achieve QKD over 400 km at around kbit/s.

In conclusion, we demonstrated a DPS-QKD experiment with a channel loss of over 72 dB using a 336 km DSF and SSPDs with a DCR of 0.01 cps. The ultralow DCR of the SSPD was achieved by introducing a cold optical filter, which completely blocks the blackbody radiation at room temperature except for signal passband.

Table 1. Results of DPS-QKD experiments

| Channel loss (dB) | DCR (cps) | Sifted key rate (bit/s) | QBER (%) | Secure key rate (bit/s) |
|---|---|---|---|---|
| 52.7 (attenuator) | 0.04 | 31.95 | 1.02 | 12.5 |
| 66 (306 km DSF) | 0.04 | $0.98^{+0.09}_{-0.12}$ | $2.64^{+0.63}_{-0.39}$ | $0.17^{+0.04}_{-0.07}$ |
| 72 (336 km DSF) | 0.01 | $0.22^{+0.10}_{-0.11}$ | $2.93^{+0.72}_{-1.18}$ | $0.03^{+0.04}_{-0.02}$ |

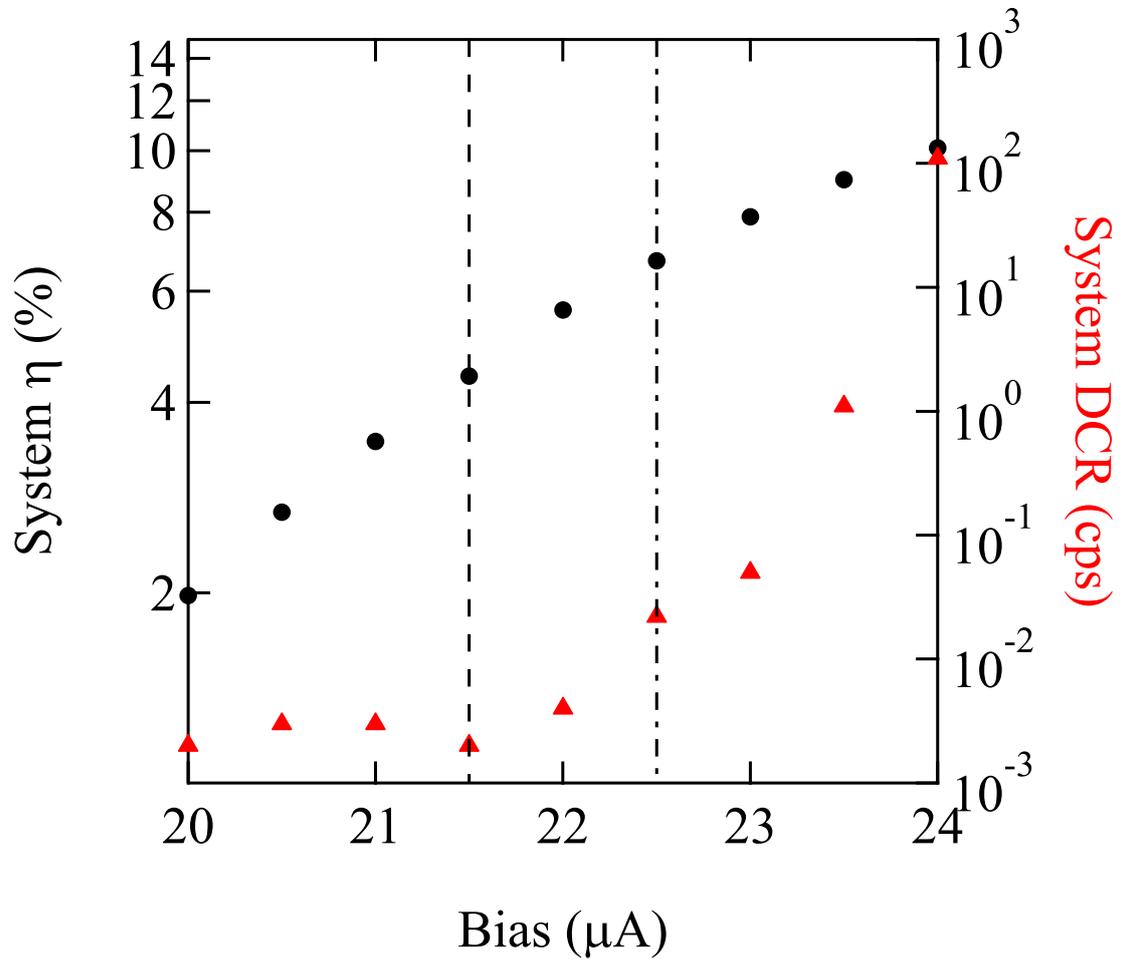

Fig. 1. Bias current dependence of the system detection efficiency (η) and the system dark count rate (DCR) at 0.38 K. The vertical line shows the bias point of the SSPD for DCR = 0.01 cps (dashed line) and 0.04 cps (dot & dashed line), respectively.

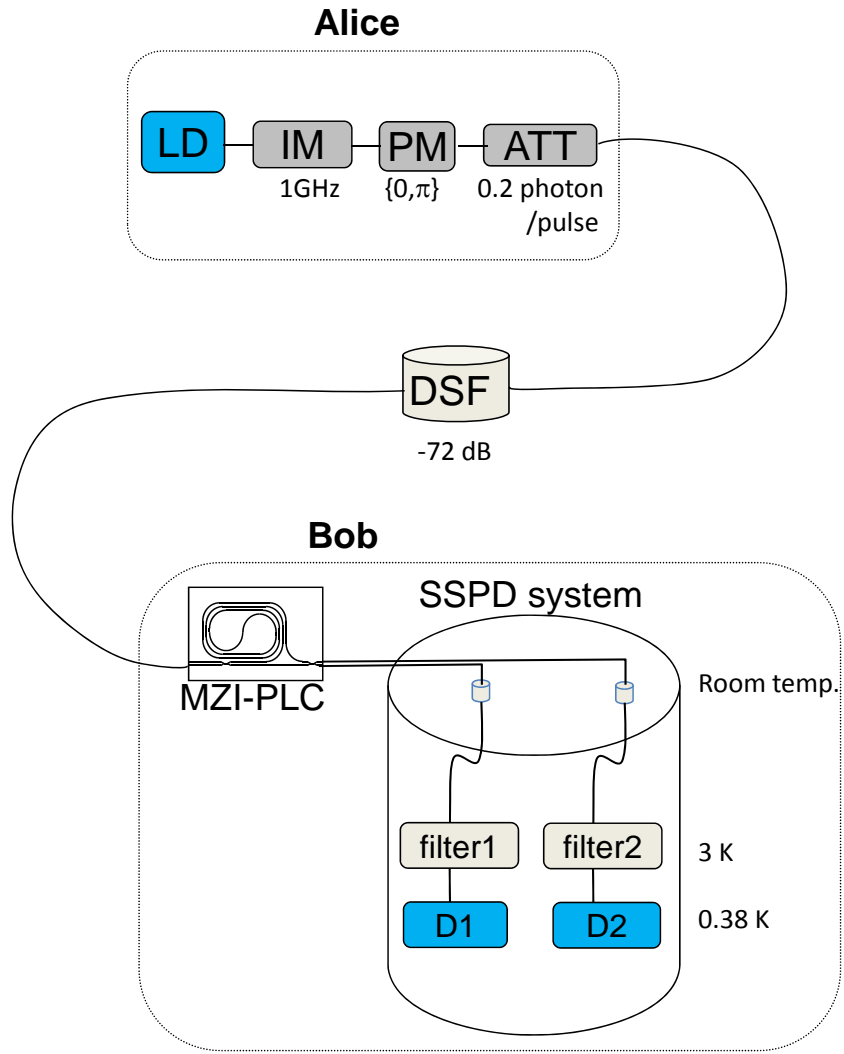

Fig. 2. Schematic diagram of DPS-QKD setup. The abbreviations are LD (laser diode), IM (intensity modulator), PM (phase modulator), ATT (attenuator), DSF (dispersion shifted fiber), MZI-PLC (1-bit delayed Mach-Zehnder interferometer based on a planar lightwave circuit), and D (SSPD device). Two SSPD devices are cooled at 0.38 K and the optical bandpass filters are cooled at 3 K in the ³He cryocooler. The blackbody radiation at room temperature (mainly near the hermetic seal of the cryocooler) is blocked at the filters cooled to 3 K except for the signal passband.

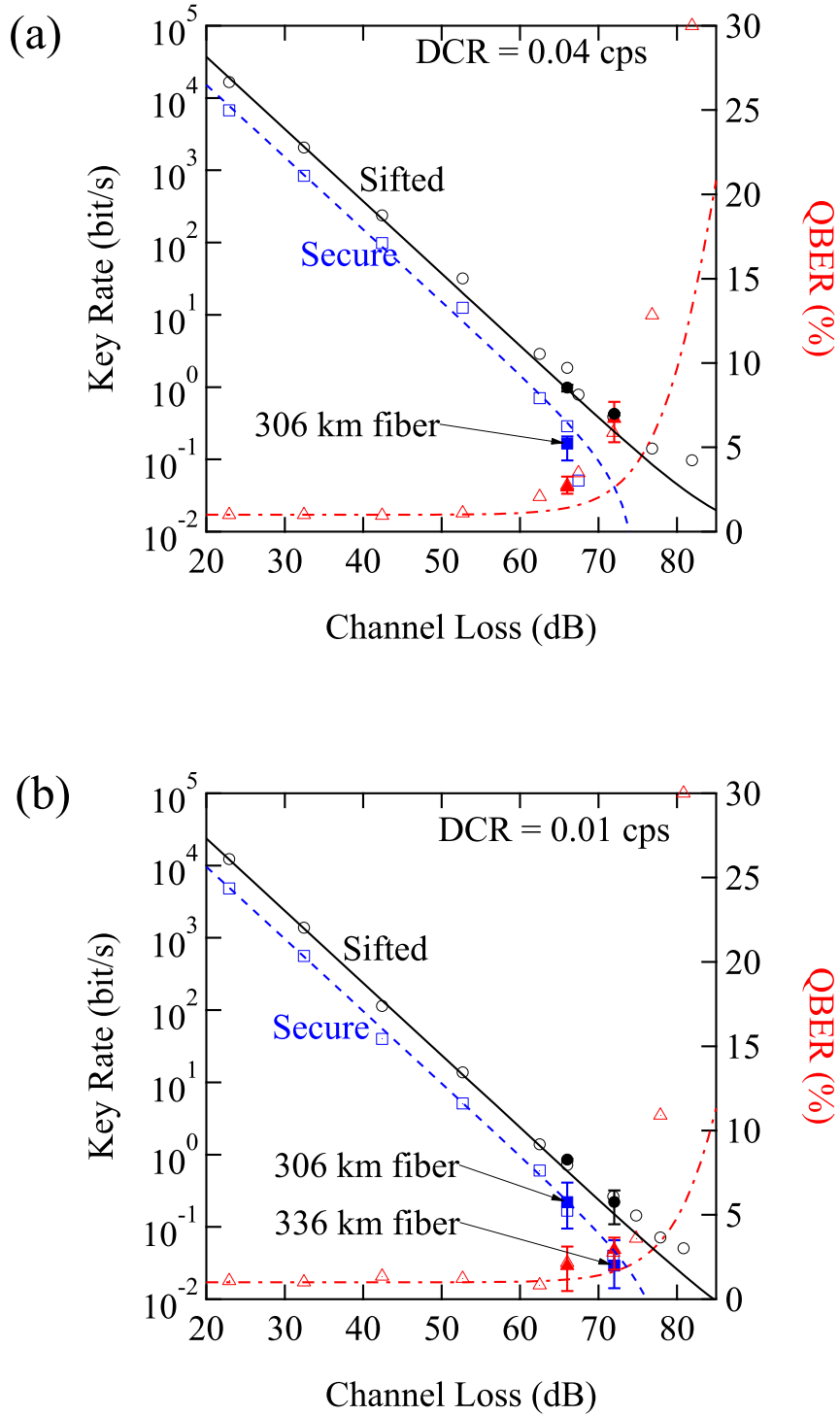

Fig. 3. Channel loss dependence of sifted key rate (circle), secure key rate (square) and QBER (triangle) with (a) DCR = 0.04 cps and (b) 0.01 cps. The open and filled symbols denote optical attenuation and DSF transmission, respectively. The secure key generation rate is calculated from the experimental results obtained for the sifted key rate and QBER. The curves are the results of theoretical fitting using Eq. (1)-(4). The channel loss does not include the MZI-PLC loss.